\documentclass[twocolumn,showpacs,aps,prd,superscriptaddress,floatfix]{revtex4-2}
\usepackage[english]{babel}
\usepackage[utf8]{inputenc}
\usepackage[displaymath, mathlines]{lineno}

\usepackage{graphicx}
\usepackage{dcolumn}
\usepackage{epsfig}
\usepackage{subfloat}
\usepackage{colordvi}
\usepackage{hhline}
\usepackage{multirow}
\usepackage{subfigure}
\usepackage{rotating}
\usepackage{hyperref}
\usepackage{xcolor}
\usepackage{url}
\usepackage{xcolor}
\usepackage[left=20mm,right=13mm,top=35mm,columnsep=20pt]{geometry} 
\usepackage{adjustbox}
\usepackage{placeins}
\usepackage[T1]{fontenc}
\usepackage{lipsum}
\usepackage{csquotes}
\usepackage{siunitx}
\usepackage{textcomp}
\usepackage{ragged2e}

\usepackage{multirow}

\usepackage{titlesec}

\usepackage{tabularx}

\addto\captionsenglish{}
\addto\captionsenglish{}

\begin{document}
\title{Precision measurement of the branching fraction of \boldmath $J/\psi\rightarrow K^+K^-$ via  $\psi(2S)\rightarrow \pi^+\pi^-J/\psi$}

  \author{ 
M.~Ablikim$^{1}$, M.~N.~Achasov$^{4,b}$, P.~Adlarson$^{75}$, X.~C.~Ai$^{80}$, R.~Aliberti$^{35}$, A.~Amoroso$^{74A,74C}$, M.~R.~An$^{39}$, Q.~An$^{71,58}$, Y.~Bai$^{57}$, O.~Bakina$^{36}$
% ...
\vspace{0.2cm}\\
(BESIII Collaboration)
\vspace{0.2cm}\\
{\it
$^{1}$ Institute of High Energy Physics, Beijing 100049, People's Republic of China\\
% ...
$^{4}$ Budker Institute of Nuclear Physics SB RAS (BINP), Novosibirsk 630090, Russia\\
% ...
$^{35}$ Johannes Gutenberg University of Mainz, Johann-Joachim-Becher-Weg 45, D-55099 Mainz, Germany\\
$^{36}$ Joint Institute for Nuclear Research, 141980 Dubna, Moscow region, Russia\\
% ...
$^{39}$ Liaoning Normal University, Dalian 116029, People's Republic of China\\
% ...
$^{57}$ Southeast University, Nanjing 211100, People's Republic of China\\
$^{58}$ State Key Laboratory of Particle Detection and Electronics, Beijing 100049, Hefei 230026, People's Republic of China\\
% ...
$^{71}$ University of Science and Technology of China, Hefei 230026, People's Republic of China\\
% ...
$^{74}$ University of Turin and INFN, (A)University of Turin, I-10125, Turin, Italy; (B)University of Eastern Piedmont, I-15121, Alessandria, Italy; (C)INFN, I-10125, Turin, Italy\\
$^{75}$ Uppsala University, Box 516, SE-75120 Uppsala, Sweden\\
% ...
$^{80}$ Zhengzhou University, Zhengzhou 450001, People's Republic of China
\vspace{0.2cm}\\
% ...
$^{b}$ Also at the Novosibirsk State University, Novosibirsk, 630090, Russia\\
% ...
}

}

\email[Electronic address: ]{besiii-publications@ihep.ac.cn}
\date{\today} 

\begin{abstract}
 Using a sample of 
$448.1 \times 10^6$ $\psi(2S)$
 events collected with the BESIII detector, we perform a study of the decay $J/\psi\rightarrow K^+K^-$  via  $\psi(2S)\rightarrow \pi^+\pi^-J/\psi$.
 The branching fraction of $J/\psi\rightarrow K^+K^-$ is determined to be $\mathcal{B}_{K^+K^-}=(3.072\pm 0.023({\rm stat.})\pm 0.050({\rm syst.}))\times 10^{-4}$, which is consistent with 
 previous measurements
 but with significantly improved precision.    
\end{abstract}


\maketitle

\section{Introduction} \label{intro} The decay of narrow vector states
of charmonium, like $J/\psi$, into states of light quarks can proceed
via $c\bar{c}$ annihilation into a virtual photon or three gluons. In
the case of the decay to two kaons, regardless of the charge of the
kaons, both strong and electromagnetic interactions contribute with
comparable strength, and the branching fraction (BF) depends
critically on their relative phase~\cite{Czyz}.  Neglecting the
interference between the continuum and resonance amplitudes may affect
the measurement of the BF, $\mathcal{B}_{K^+K^-}$, in the reaction
$e^+e^- \rightarrow K^+K^-$.

Early phenomenological studies argued that this relative phase is
quite large, close to $\pm 90^{\circ}$~\cite{Suzuki,Rosner,Suzuki2,Lopez}. The contribution of the continuum has been investigated also in Refs.~\cite{Mo,Wang:2002np}, showing that the interference between continuum and resonance amplitudes may affect the $J/\psi$ decays. Furthermore, the authors show the crucial role in confirming the large universal phase between the strong and the electromagnetic amplitudes, which can affect precision measurements of the branching ratio of charmonia decays with special focus on the decays in two pseudoscalars.
In Ref.~\cite{Czyz}, it is demonstrated that the direct coupling of $J/\psi$ to the hadronic final state is important in the $K^+K^-$ case and a combination of measurements on and off resonance is essential to obtain the relative magnitude and phase of quantum electrodynamics (QED) and hadronic amplitudes.

The most recent $\mathcal{B}_{K^+K^-}$ measurement was reported by BaBar~\cite{babar}, using the untagged initial state radiation (ISR) technique, to be $\mathcal{B}_{K^+K^-}=(3.36 \pm 0.20({\rm stat.}) \pm 0.12({\rm syst.}) \times 10^{-4}$. They applied a correction for the effect of the interference with the continuum, after the determination of the relative phase, by combining their result and the ones from other experiments for the $J/\psi\to K^0_SK^0_L$ decay~\cite{ksklbes,cleoc}.
By using the result from Ref.~\cite{ksklbes}, the BaBar collaboration determined a relative phase $\varphi=(97\pm5)^{\circ}$ or $(-97\pm 5)^{\circ}$;  on the other hand, by  using the results in Ref.~\cite{cleoc}, the phase was determined to be $\varphi=(111\pm 5)^{\circ}$ or $(-109\pm 5)^{\circ}$.   Using the latter determination of $\varphi$, $\mathcal{B}_{K^+K^-}$ was measured to be $(3.22 \pm 0.20({\rm stat.}) \pm 0.12({\rm syst.})) \times 10^{-4}$ or $(3.50 \pm 0.20({\rm stat.}) \pm 0.12({\rm syst.}) )\times 10^{-4}$, depending on the sign of the phase. It is worthwhile to notice that the shifts due to the interference are large, {\it i.e.} about 5$\%$, and should be taken into account in the $e^+e^- \rightarrow K^+K^-$ reaction for precision measurements.

The study of this decay channel, by means of  $\psi (2S)\rightarrow \pi^+\pi^-J/\psi$, allows us to measure $\mathcal{B}_{K^+K^-}$ with no need to take into account the interference of the resonant amplitude with the continuum. It is advantageous, being the most probable $\psi (2S)$ decay ($\mathcal{B}(\psi (2S)\rightarrow \pi^+\pi^-J/\psi)=(34.98 \pm 0.02 \pm 0.45)\%$~\cite{BES_ppj}), and the most accessible experimentally. 
For studies of  $J/\psi$ decays, the $J/\psi$ events can be tagged by the recoiling mass against the $\pi^+\pi^-$ system in the decay $\psi(2S)\rightarrow \pi^+\pi^-J/\psi$. This $J/\psi$ sample is automatically free of any contamination of $K^+K^-$, $p\overline{p}$, $\mu^+\mu^-$ and $e^+e^-$ produced directly at the center of mass energy ({\rm cme}) of $\sqrt{s}= 3.097$ GeV. The continuum contribution is negligible in the subsequent $J/\psi$ decay, making this sample relatively clean and simple to analyze.
The most recent measurement of $\mathcal{B}_{K^+K^-}$ in this channel was performed by Metreveli {\it et al.} using CLEO-c data~\cite{cleoc}, with a result of $(2.86 \pm 0.09 \pm 0.19) \times 10^{-4}$; this measurement currently is the only one determining the world average by the Particle Data Group (PDG)~\cite{pdg22}. In the same article the relative phase in $J/\psi$ decays into pseudoscalar pairs was reported to be $\varphi= (73.5^{+5.0}_{-4.5}) ^{\circ}$, obtained by means of the BFs of $J/\psi \to K^+K^-$, $J/\psi \to \pi^+\pi^-$ and $J/\psi \to K^0_SK^0_L$.

A sample of 448.1 $\times 10^6$ $\psi(2S)$ events collected in 2009 and 2012 with the BESIII detector offers, by virtue of being about 20 times larger than the CLEO-c statistics, a unique opportunity to improve the precision of $\mathcal B_{K^+K^-}$. 
It is worthwhile to notice that the sign of $\sin \varphi$ for $J/\psi$ decays can be determined experimentally from the relative difference, $\delta \mathcal{B}_{K^+K^-}/ \mathcal{B}_{K^+K^-}$, between precise measurements of BFs in both $e^+e^- \rightarrow K^+K^-$ reaction and $\psi(2S)\rightarrow \pi^+\pi^-J/\psi$, as suggested in Ref.~\cite{babar}.

This paper describes the measurement of $\mathcal{B}_{K^+K^-}$ via $\psi(2S)\rightarrow \pi^+\pi^-J/\psi$ with BESIII data and it is organized as follows.
Section~\ref{samples} contains a brief description of the BESIII detector and data and Monte Carlo (MC) samples.
In Sec.~\ref{method} we describe the analysis strategy, that foresees a measurement relative to the precisely measured BF of the $J/\psi\to\mu^+\mu^-$ decay.
In Sec.~\ref{selections} we present the event selection  and in Sec.~\ref{bkg} the background evaluation can be found. The $\mathcal{B}_{K^+K^-}$ measurement is presented in Sec.~\ref{misura} while the  systematic uncertainties are discussed in Sec.~\ref{sys}. Section~\ref{summary}, with summary and conclusions, closes the paper.

\section{BESIII detector, data and MC samples}
\label{samples}
In this analysis we use two large data samples of $\psi(2S)$ decays, collected with the BESIII detector in 2009 and in 2012. The numbers of $\psi(2S)$ events were determined to be $(107.0 \pm 0.8) \times 10^6$ and $(341.1 \pm 2.1) \times 10^6$ \cite{Npsi}, respectively.
The BESIII detector records symmetric $e^+e^-$ collisions 
provided by the BEPCII storage ring~\cite{Yu:IPAC2016-TUYA01}, which operates in the {\rm cme} range from 2.00 to 4.95~GeV.
BESIII has collected large data samples in this energy region~\cite{Ablikim:2019hff}. The cylindrical core of the BESIII detector covers 93\% of the full solid angle and consists of a helium-based
 multi-layer drift chamber~(MDC), a plastic scintillator time-of-flight
system~(TOF), and a CsI(Tl) electromagnetic calorimeter~(EMC),
which are all enclosed in a superconducting solenoidal magnet
providing a 1.0~T (0.9~T in
2012) magnetic field. The solenoid is supported by an
octagonal flux-return yoke with resistive plate counter muon
identification modules interleaved with steel. 

The charged-particle momentum resolution at $1~{\rm GeV}/c$ is
$0.5\%$, and the 
resolution of the specific energy loss, ${\rm d}E/{\rm d}x$,
is $6\%$ for electrons
from Bhabha scattering. The EMC measures photon energies with a
resolution of $2.5\%$ ($5\%$) at $1$~GeV in the barrel (end cap)
region. The time resolution in the TOF barrel region is 68~ps, while
that in the end cap region is 110~ps. Details about the design and performance of the
BESIII detector are given in Ref.~\cite{Ablikim:2009aa}. 

Monte Carlo simulated data samples produced with a {\sc
geant4}-based~\cite{geant4} software package, which
includes the geometric description of the BESIII detector~\cite{detector2} and the
detector response, are used to determine detection efficiencies, to estimate backgrounds and for event selection optimization. The simulation models the beam
energy spread and ISR in the $e^+e^-$
annihilations with the generator {\sc
  kkmc}~\cite{ref:kkmc,ref:kkmc1}. 
An `inclusive' MC (INC-MC) sample includes the production of the
$\psi(2S)$ resonance, the ISR production of $J/\psi$, and
the continuum processes incorporated in {\sc
kkmc}~\cite{ref:kkmc,ref:kkmc1,Ping:2008zz}.
All particle decays are modeled with {\sc
evtgen}~\cite{ref:evtgen,Ping:2008zz} using BFs 
either taken from the
PDG~\cite{pdg}, when available,
or otherwise estimated with {\sc lundcharm}~\cite{ref:lundcharm,ref:lundcharm1}.
Final state radiation (FSR) from charged final state particles is incorporated using the {\sc
  photos} package~\cite{photos}.
The INC-MC samples correspond to 106 millions $\psi (2S)$ events for 2009 and 400 millions $\psi (2S)$ events for 2012.\par
To evaluate the detection efficiency, exclusive signal MC simulations are performed with equivalent sizes much larger than the expected data statistics. The $\psi(2S)$ production has been previously described and the subsequent decays are generated using {\sc evtgen}~\cite{ref:evtgen,Ping:2008zz} with the  {\sc JPIPI} module for $\psi(2S)\rightarrow \pi^+\pi^- J/\psi$, {\sc VSS} for $ J/\psi \rightarrow K^+K^-$  and  {\sc VLL} for $J/\psi \rightarrow \mu^+\mu^-$ .

\section {Method}
\label{method}
In order to reduce the experimental systematic uncertainties, the BF of $J/\psi \rightarrow K^+K^-$ is determined relative to the well known $J/\psi \rightarrow \mu^+\mu^-$ channel. The strategy for this analysis is to measure the ratio 
\begin{equation}
\mathcal{R}\equiv \frac{\mathcal{B}_{K^+K^-}}{\mathcal{B}_{\mu^+\mu^-}}.
\end{equation}
This ratio can be re-written as
\begin{equation}
\mathcal{R}=\frac{\mathcal{B}_{K^+K^-}}{\mathcal{B}_{\mu^+\mu^-}}=\frac{\frac{N_{K^+K^-}}{\epsilon_{K^+K^-}\times N_{\psi(2S)}}}{\frac{N_{\mu^+\mu^-}}{\epsilon_{\mu^+\mu^-}\times N_{\psi(2S)}}}=\frac{N_{K^+K^-}\\
\times \epsilon_{\mu^+\mu^-}}{N_{\mu^+\mu^-} \times \epsilon_{K^+K^-}},
\label{eq:R}
\end{equation}
where $N_{K^+K^-}$ and $N_{\mu^+\mu^-} $ are the observed numbers of events of $J/\psi \to K^+K^-$ and $J/\psi \to \mu^+\mu^-$, respectively, while $\epsilon_{K^+K^-}$ and $\epsilon_{\mu^+\mu^-}$ are the corresponding reconstruction efficiencies.
In this way, the systematic uncertainties in the total number of $\psi(2S)$ events and the tracking efficiencies of charged particles, are canceled in the $\mathcal{R}$ measurement.
Combining the obtained $\mathcal{ R}$ with the precise world average of $\mathcal{B}_{\mu^+\mu^-}$ from the PDG~\cite{pdg}, $(5.916\pm0.033)\%$, $\mathcal{B}_{K^+K^-}$ can be measured.

\section{Event selection}
\label{selections}
The channel we are investigating is $\psi(2S)\rightarrow \pi^+\pi^-J/\psi$  with the subsequent decay $J/\psi \rightarrow K^+K^-$.

The total number of charged tracks in an event must be four with net charge equal to zero. 
Charged tracks are reconstructed using the MDC. For each track, the point of closest approach to the interaction point must be within 1 cm in the plane perpendicular to the beam direction and within 10 cm along the beam direction. Moreover, to guarantee better agreement between data and MC simulation, we require them to have a polar angle in the range $-0.80<\cos \theta < 0.80$, where $\theta$ is defined with respect to the $z$-axis, taken to be the symmetry axis of the MDC.
Furthermore, a vertex fit in which all charged tracks are forced to originate from a common vertex is performed and required to be successful. The parameters of daughter tracks after the vertex fit are updated.

We require two pion candidates and two kaon candidates, both oppositely charged, in each event. No particle identification is required to avoid additional systematic uncertainty.
The two tracks with lower momentum (less than 1.0 GeV/$c$) are assigned as the pion candidates, as expected by the exclusive MC sample. To veto the background events associated with photon conversion we require  the cosine of the opening angle 
between the two pions, $\cos \theta_{\pi\pi}$, to be less than 0.5.
The recoiling mass against the two selected pions is calculated as
\begin{equation}
\rm{RM}({\pi^+\pi^-})=\sqrt{(p_{\rm{cme}} - p_{\pi^+} - p_{\pi^-})^2}
\end{equation}
where $p_{\rm{cme}}$ and $p_{\pi^{\pm}}$ are the four-momenta in the center of mass system and for the two charged pion candidates, respectively.
\begin{figure}[htb!!]
\centering
\includegraphics[width=0.49\textwidth]{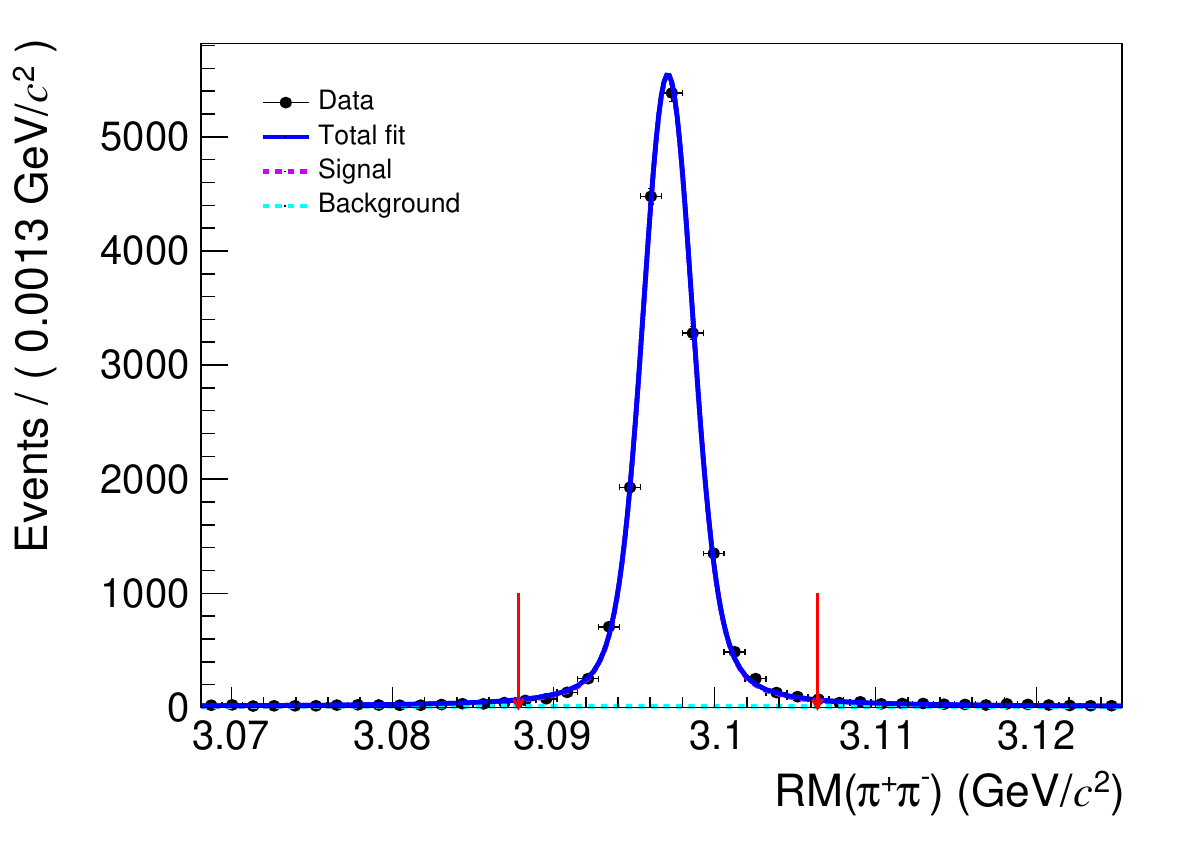}
\caption{Distribution of \rm{RM}$(\pi^+\pi^-)$ against the ($\pi^+\pi^-$) system of the selected candidates. The best fit of a Voigt function to the data is shown as the blue line. The red arrows show the $\pm \,5\sigma_v$ range. }
\label{fig:recoil1}
\end{figure}
The recoiling mass ($\rm{RM(}{\pi^+\pi^-})$) distribution of the accepted candidates is shown in Fig.~\ref{fig:recoil1}, dominated by the production of the $J/\psi$ resonance.
To tag the $J/\psi$ production, the RM$(\pi^+\pi^-)$ range between 3.087 and 3.107 GeV/$c^2$ is required, corresponding to about $\pm 5\,\sigma _v$, where the width $\sigma _v$ is obtained by fitting the {${\rm RM}({\pi^+\pi^-})$ distribution with a Voigt function.
The remaining two good charged particles can be kaon candidates, for which we require their momenta greater than 1.2 GeV/$c$, as suggested by the exclusive signal MC sample.
The opening angle $\theta_{K^+K^-}$ between the two decay products is calculated in the reference frame of the $J/\psi$, requiring $\cos \theta_{K^+K^-} < -0.95$.
The yields of the $J/\psi \to e^+e^-$ and $J/\psi \to \mu^+\mu^-$ overwhelm the $J/\psi \to K^+K^-$ one, since their BFs are several orders of magnitude larger.
To veto the background of $J/\psi \to e^+e^-$, an efficient selection can be found using the ratio $E_\text{dep}/p$, in which $E_\text{dep}$ is the energy deposited in the EMC and $p$ is the reconstructed momentum in the MDC for a decay product of $J/\psi$, requiring it to be less than $0.8/c$.
Further selection is applied on the total energy deposited in the EMC for an event, $E_{\rm EMC}$, which is required to be between 0.3 and 2.5 GeV.
To perform the measurement of $\mathcal{B}_{K^+K^-}$, we calculate the scaled visible energy ($X_{\rm{vis}}$) of the two particles in the decay final state, which are $ K^+$ and $K^-$ candidates, as
\begin{equation}
X_{\rm{vis}}= \frac{E_{K^+}+E_{K^-}}{E_{J/\psi}},
\label{eq:evis}
\end{equation}
where $E_{K^{\pm}}$ are the energies of the kaon candidates calculated using the kaon mass hypothesis in the center-of-mass frame and $E_{J/\psi} $ is the $J/\psi$ energy, in the same reference frame, corresponding to the $J/\psi$ rest mass.
The $J/\psi \to K^+K^-$ signal is expected to center around $X_{\rm{vis}}=1.0$, while at lower values two final state particles with higher mass can be found, mainly from $J/\psi \to p\bar p$ decay due to the lack of PID requirements, and at higher values the ones with lower mass (mainly $J/\psi \to \mu^+\mu^-$).
We define two different intervals in $X_{\rm{vis}}$ to tag the two decays of interest: the $J/\psi\to K^+K^-$ signal region (KKR) and the $J/\psi\to \mu^+\mu^-$ signal region (MMR), between 0.98 and 1.01 and between 1.02 and 1.07, respectively.
The asymmetric range of KKR is chosen because of the significant tail of the $J/\psi \to \mu^+\mu^-$ distribution, shown in Fig.~\ref{fig:allevis}, where the MC distributions are shown for INC-MC with MC-truth matching for the $J/\psi \to K^+K^-$ and $J/\psi \to \mu^+\mu^-$ decays. 
\begin{figure}[htb!!]
\centering
\includegraphics[width=0.49\textwidth]{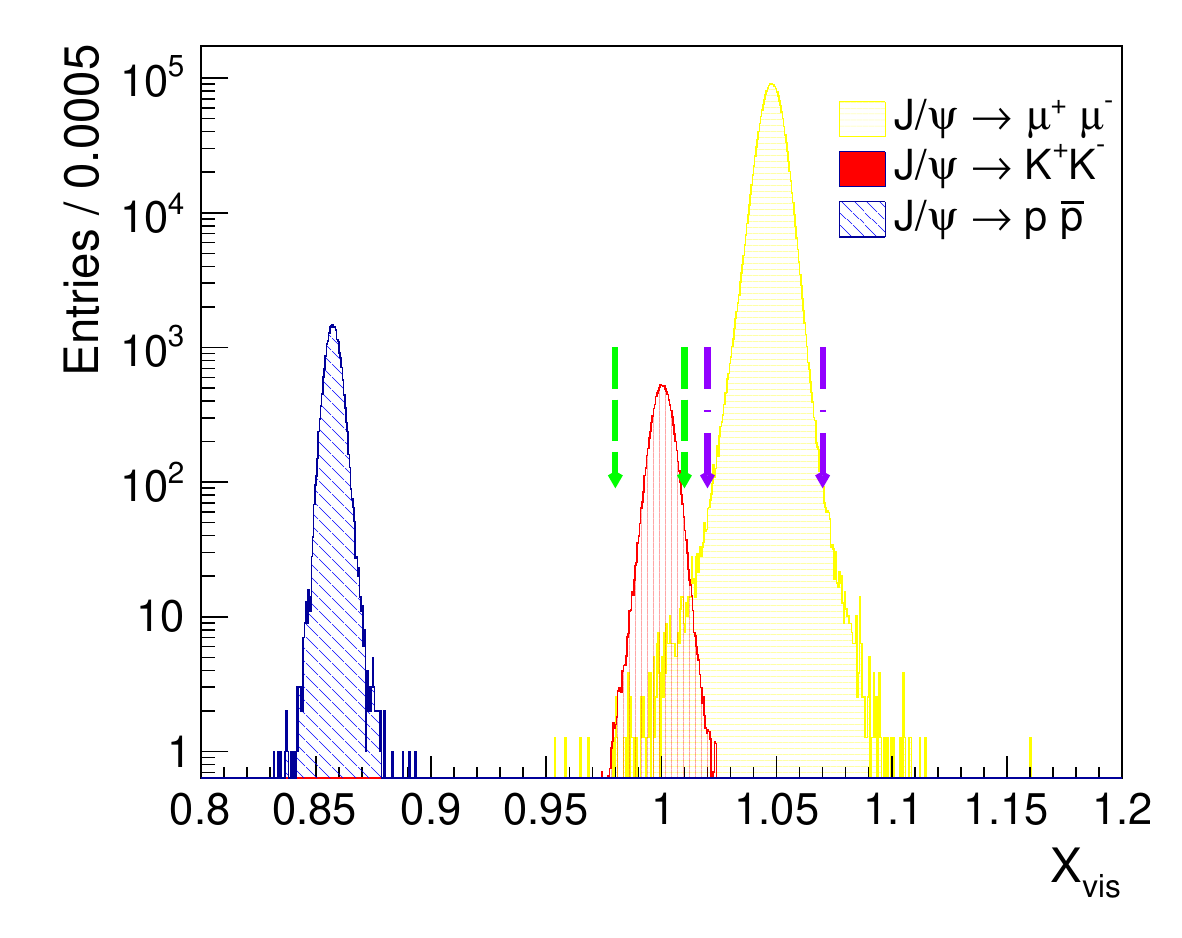}
\caption{Distribution of $X_{\rm vis}$, for the INC-MC sample, with MC-truth matching applied, in log-scale, pattern-filled with horizontal lines in red for $J/\psi\to K^+K^-$, with vertical lines in yellow for $J/\psi\to \mu^+\mu^-$ and with oblique lines in blue for $J/\psi\to p \bar{p}$. The green dashed line arrows show the KKR and the purple dot dashed line ones the MMR, as described in the text. The normalization is arbitrary.}
\label{fig:allevis}
\end{figure}
Momentum conservation in the reconstructed $J/\psi$ decays is required by demanding, in the decay, the magnitude of the normalized vector sum of the total momentum, $|\sum_i {\bf p_i}|/E_{J/\psi}$, to be less than 0.025. 
This
allows us to reject most of the background from decays to more than two bodies.\par
\begin{figure}[htb!!]
\centering
\includegraphics[width=0.45\textwidth]{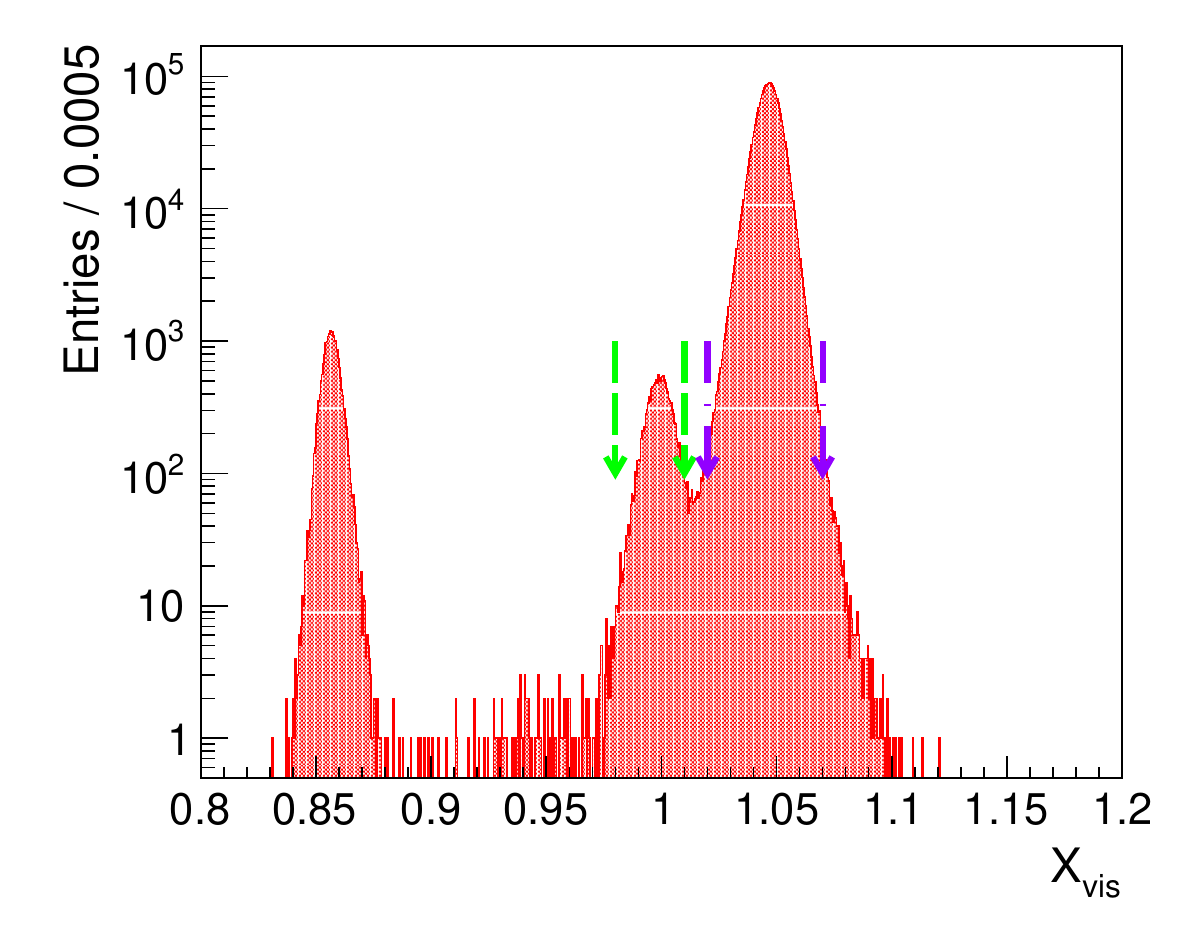}
\caption{Distribution of $X_{\rm vis}$ for data, in log-scale. The green dashed line arrows show the KKR and the purple dot dashed line ones the MMR, as described in the text. 
}
\label{fig:dataevis}
\end{figure}
After all these requirements the $X_{\rm{vis}}$ distribution features three structures, as shown in Fig.~\ref{fig:dataevis}. The one centering around $X_{\rm{vis}}=0.85$ is well separated, and consists mainly of the residual $J/\psi \to p\bar p$ decays. A further selection is applied, requiring the $X_{\rm vis}$ to be greater than 0.90. The third structure, centering around $X_{\rm{vis}}=1.05$, is well separated, but the tail in the KKR must be evaluated in the fit procedure.
\section{Background estimation}
\label{bkg} 
We perform a background study using the INC-MC samples, described in Sec.~\ref{samples}, thanks to the MC-truth matching and to the TopoAna package~\cite{topo-ana}. After all the selections, except the $X_{\rm{vis}}$ ones, described in Sec.~\ref{selections}, the main background components are shown in Fig.~\ref{fig:bkgevis2012}.
In MMR, the background is mainly due to $\psi(2S)\rightarrow \pi^+\pi^- J/\psi, J/\psi\rightarrow e^+e^-/\pi^+\pi^-$, accounting for about 0.67\%, as shown in the bottom of Fig.~\ref{fig:bkgevis2012}. Non-$\pi^+\pi^- J/\psi$ background is negligible.
In KKR, the background fraction is about 6\%. It is dominated by non-$\pi^+\pi^- J/\psi$ events that account for about 70\% of the total background, while about 30\% is due to the tail of $J/\psi \to \mu^+\mu^-$, dominating the background involving  $J/\psi$ production as shown in Fig.~\ref{fig:bkgevis2012}.  This background can be subtracted in the fit procedure.  The non-$\pi^+\pi^- J/\psi$ background is a peaking background, mainly involving the $K_1(1270)^{\pm}$ resonance production. The main background channel (charge conjugation is implied) is the decay $\psi(2S)\to K^{-} K_1(1270)^{+}$ with the subsequent decays $K_1(1270)^{+}\to \pi^{+ } K^{*}_0(1430)$ and $K^{*}_0(1430)\to\pi^{-} K^{+}$. Large uncertainties on the decay BFs related to these channels are reported in the PDG~\cite{pdg22}.
\begin{figure}[htb!!]\centering
  \centering
  \includegraphics[width=0.49\textwidth]{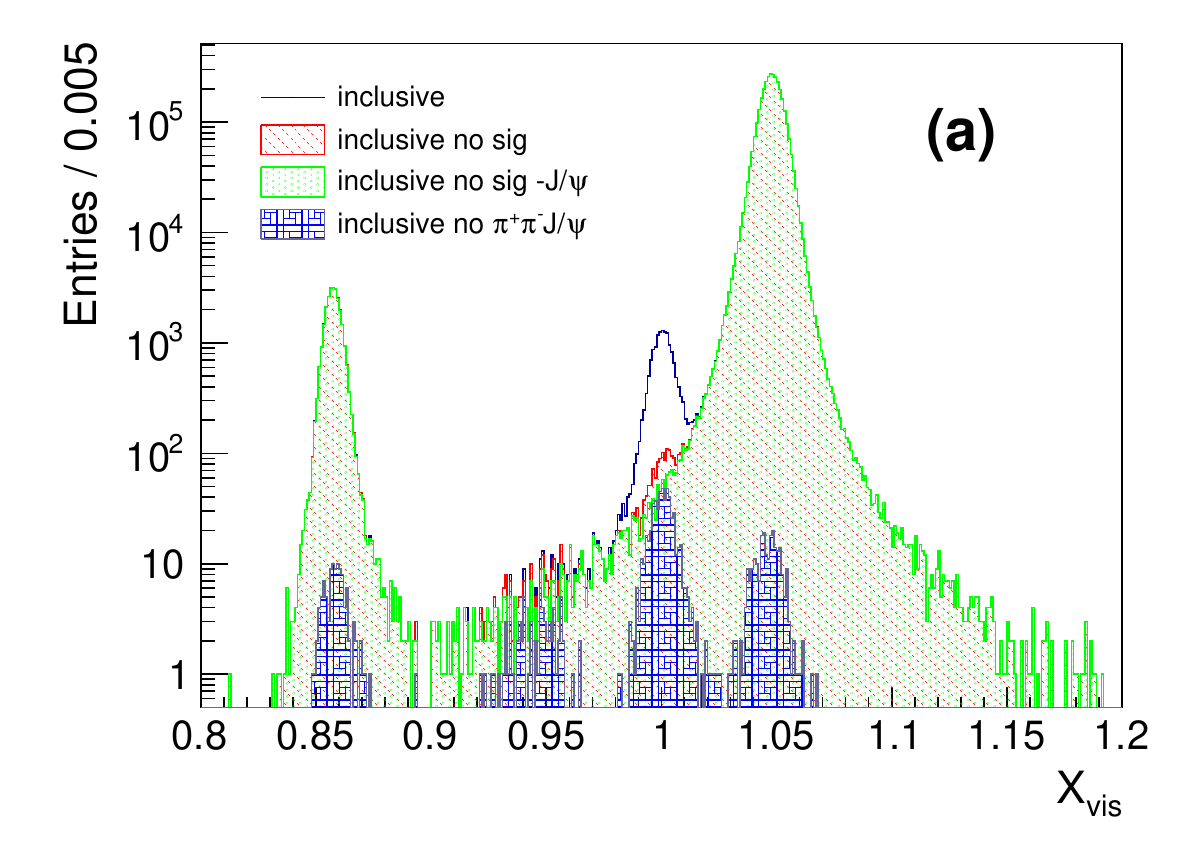}
  
\includegraphics[width=0.49\textwidth]{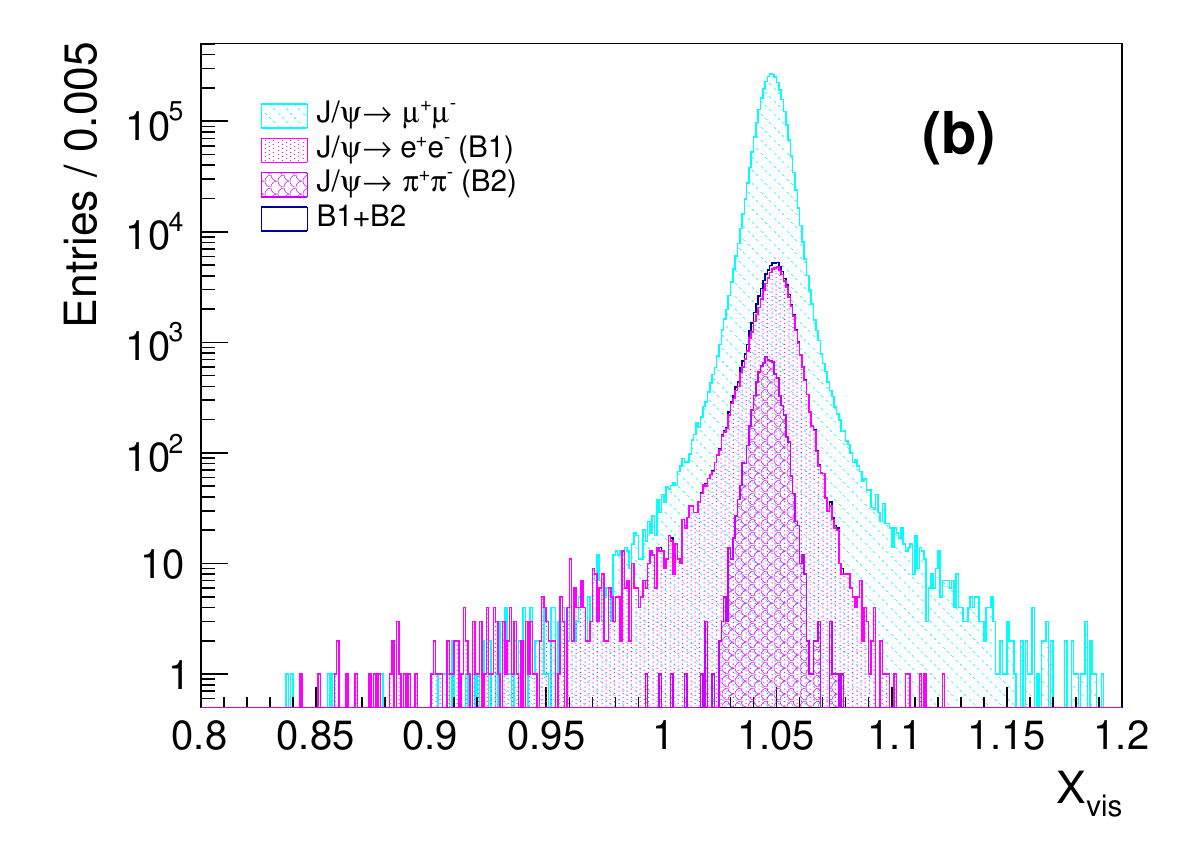}
\caption{Distributions of $X_{\rm{vis}}$ for the inclusive Monte Carlo
sample, broken down according to MC truth information. In the figure {\bf (a)}, the $X_{\rm{vis}}$
distribution for events in KKR is plotted. The blue line indicates the
total INC-MC distribution, the red histogram gives all background events
in KKR, and the green and blue histograms give background events with a
$J/\psi$ and without $\pi^+\pi^-J/\psi$, respectively. Likewise, the plot {\bf (b)}
shows the $X_{{\rm vis}}$ distribution for MMR broken down into its relevant contributions, as indicated in the legend.}
  \label {fig:bkgevis2012}
\end{figure}
\begin{figure}[htb!!]\centering 
  \includegraphics[width=0.49\textwidth]{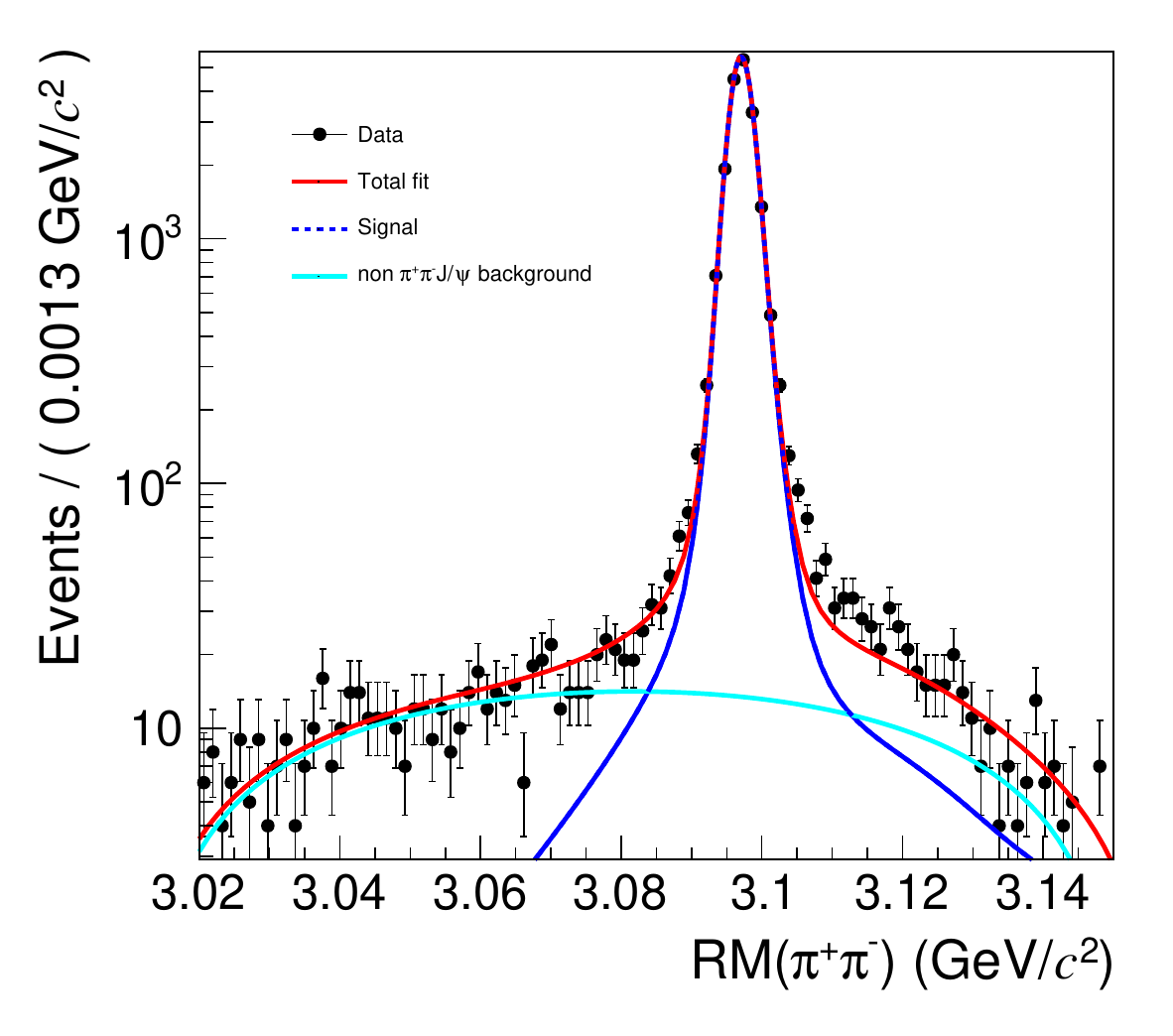}
  \caption{Distribution of RM$(\pi^+\pi^-)$ of the accepted candidates for the full data sample with the best fit result in KKR in red in logarithmic scale. The signal is shown with blue dashed line and the non-$\pi^+\pi^-  J/\psi$ background in cyan. The black points are data. }
  \label {fig:bkgrecoil2012}
\end{figure}
For this reason, the non-$\pi^+\pi^- J/\psi$ background fraction in KKR is evaluated by a data-driven method. An unbinned extended maximum likelihood fit is performed to the $\rm{RM}({\pi^+\pi^-}$) distribution in KKR in the range $[3.02-3.15]$ GeV/$c^2$. 
The signal (all $J/\psi$ decays) is modeled by the MC-simulated signal shape, convolved with a Gaussian resolution function to account for data-MC differences, and the background by a second-order Chebyshev polynomial function. The parameters of the Gaussian and all yields are left free in the fit. The integral of the best fit function in the defined signal range, [3.087, 3.107] GeV/$c^2$, allows us to evaluate the peaking background contribution, as shown in Fig.~\ref{fig:bkgrecoil2012}.
The ratio between non-$\pi^+\pi^-  J/\psi$ background and the total number of events is found to be ($1.110\pm 0.157$)\%, ($1.257 \pm 0.095$)\% and ($1.145 \pm 0.079$)\% for the 2009, 2012, and 2009+2012 (full) data samples, respectively.

\section{Measurement of the Branching Fraction of $J/\psi\rightarrow K^+K^-$}
\label{misura}
\subsubsection{Signal yields and efficiencies}
The yields of the $ J/\psi$ decays into two charged kaons and into two muons can be determined simultaneously by a fit to the $X_{\rm{vis}}$ distribution, as defined in Eq.~(\ref{eq:evis}).
In the fit, the tail of $\mu^+ \mu^-$ distribution in KKR can be evaluated and subtracted.
An unbinned extended maximum likelihood fit is performed.
The probability density functions are MC-simulated shapes, extracted from the exclusive MC samples and convolved with a single Gaussian function for the $J/\psi \to K^+K^-$ decay and a double-Gaussian function for the $J/\psi \to \mu^+\mu^-$ contribution.
 The parameters of the Gaussians and both the yields are left free in the fit.
The peaking background in KKR is described by a Gaussian function, with the parameters fixed to the best fit ones obtained from the INC-MC sample, selected by MC-truth matching. The background fraction is fixed to the ratio determined by the data-driven method as described in Sec.~\ref{bkg}. 
The peaking background in MMR is described by a single MC-simulated shape for both $J/\psi \to e^+e^-$ and $J/\psi \to \pi^+\pi^-$ decays, extracted from the INC-MC sample with MC-truth matching, as well.
The background fraction is fixed to the INC-MC value in this case.
\begin{table} \caption{ The efficiencies ($\epsilon$)  and the yields ($N$) for both $J/\psi \to \mu^+\mu^-$ and $J/\psi \to K^+K^-$ decays, indicated in the subscript. }
  \begin{center}
    \begin{tabular}{|l|l|}\hline
             \hline
      $\epsilon_{K^+K^-}$&  $ 0.3732 \pm 0.0006$\\\hline
      $\epsilon_{\mu^+\mu^-}$& $ 0.3269 \pm 0.0006$ \\\hline
      $N_{K^+K^-}$ & $18176\, \pm \,135$\\\hline
             $N_{\mu^+\mu^-}$ & $3026803\, \pm \,17409$\\\hline
             \hline
    \end{tabular}
      \end{center}
    \label{tab:ingredients}
\end{table}
In Fig.~\ref{fig:fitevis} the fit results are shown for the full data sample. 
The yields  $N_{K^+K^-}$ and $N_{\mu^+ \mu^-}$ are determined by the integral of the fit function in KKR and MMR, respectively.
An input/output check is performed by means of pseudoexperiments. We use 1000 MC samples, each with the same size of data, to check the pull distribution and no bias is found, confirming the stability of the fit procedure. 
\begin{figure}[htb!!]\centering
  \includegraphics[width=0.49\textwidth]{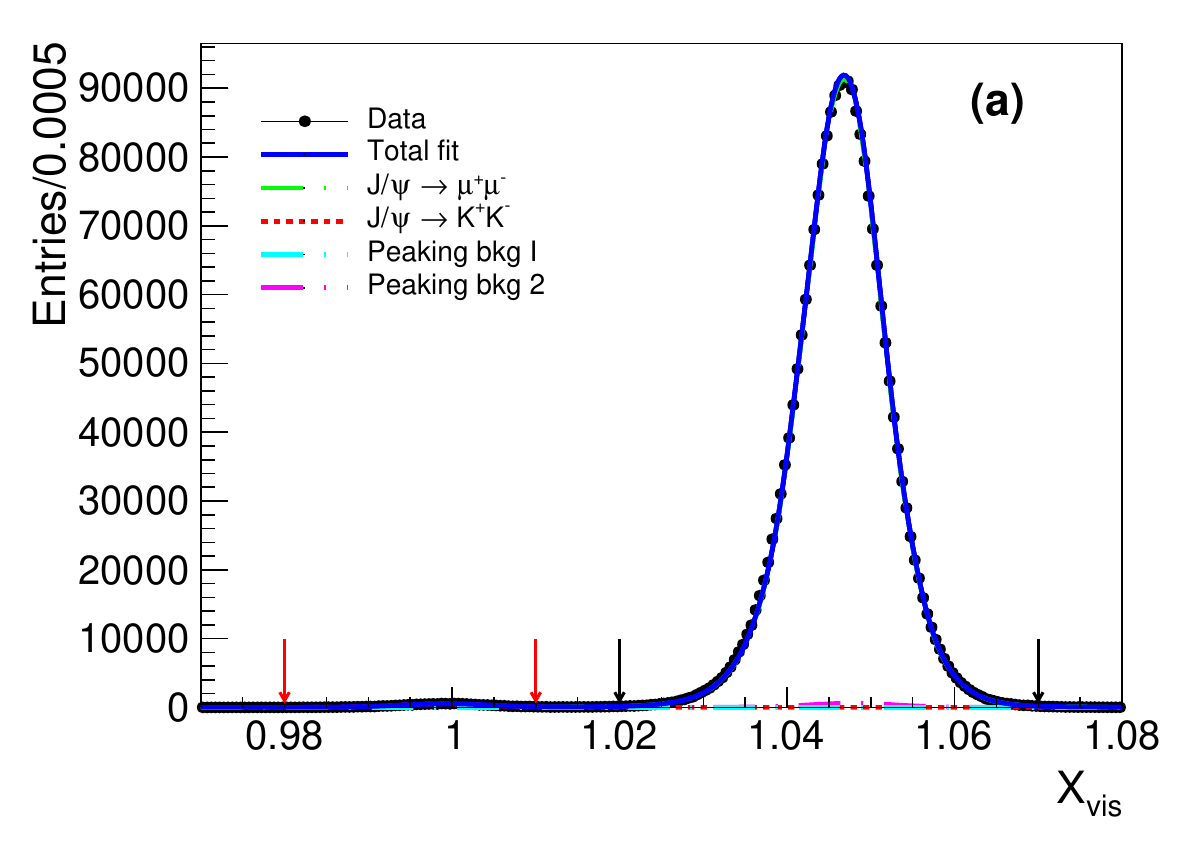}
  \includegraphics[width=0.48\textwidth]{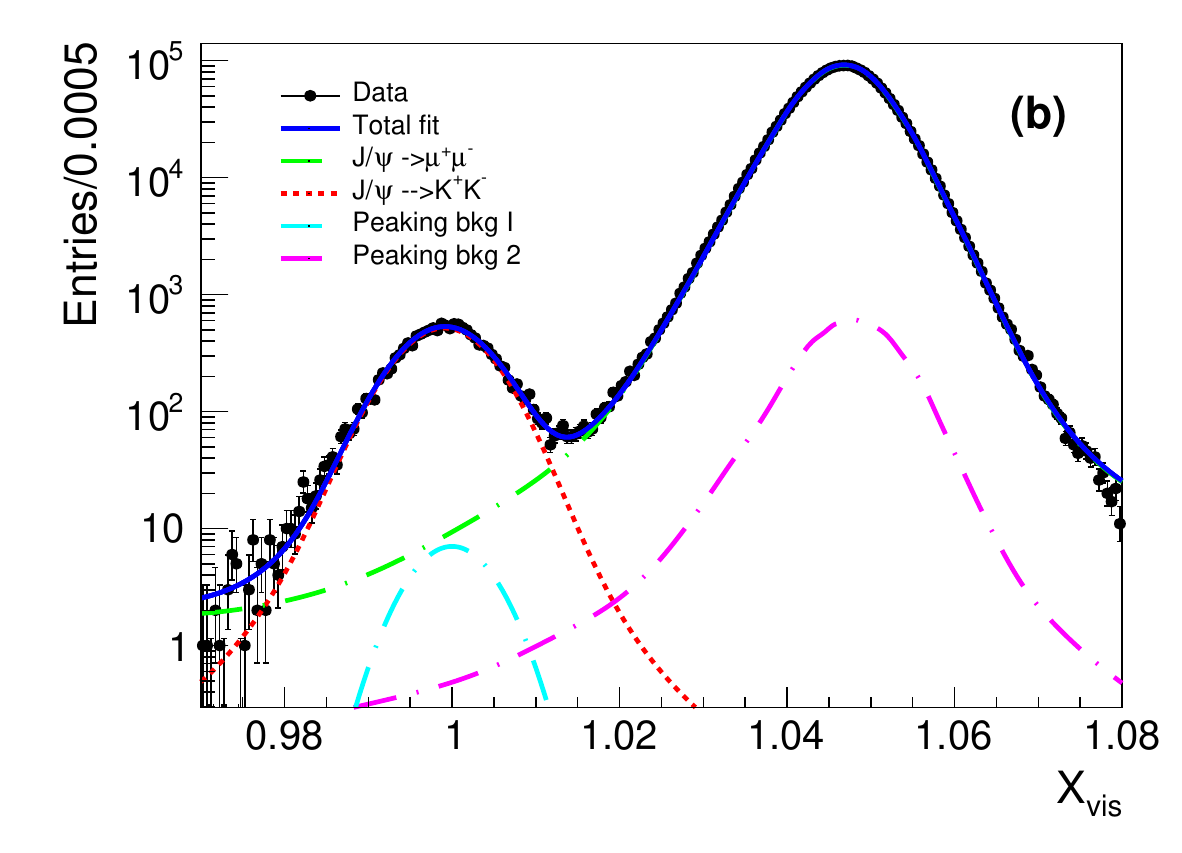}
  \caption{ Distribution of $X_{\rm vis}$ with the best fit results (blue line) in the full range in linear scale {\bf (a)}.  Plot {\bf (b)} shows the same distribution in logarithmic scale to better appreciate the fit results in KKR. The different contributions are shown by dotted curves: in red $J/\psi \to K^+K^-$, in green $J/\psi \to \mu^+\mu^-$, while in cyan and magenta the two peaking backgrounds (peaking bkg I and II) in KKR (red dashed arrows) and MMR (black dot-dashed arrows), respectively, as described in the text. The black points are data. }
  \label {fig:fitevis}
\end{figure}
The efficiencies for both $J/\psi \to K^+K^-$ and $J/\psi \to \mu^+\mu^-$ are determined by means of exclusive MC samples, described in Sec.~ \ref{samples}, separately for the two data samples. The full-sample efficiencies, determined with a weighting procedure,
are $\epsilon_{\mu^+\mu^-} = 0.3269\,\pm \, 0.0006 $ and $\epsilon_{K^+K^-}=0.3732\, \pm \,0.0006 $.

\subsubsection{ BF results}
The ratio $\mathcal{ R}  =0.005153\,\pm \,0.000041$ is determined from Eq.~(\ref{eq:R}), using the yields and efficiencies extracted so far and reported in Table~\ref{tab:ingredients}.
Using the PDG $\mathcal{B}_{\mu^+\mu^-}$ \cite{pdg}, we measure  $\mathcal{B}_{K^+K^{-}} = (3.072\pm \,0.023)\times 10^{-4}$.
The same procedure has been used for the 2012 and 2009 data samples separately with results in agreement with those of the full data sample. A further input/output check is performed using the INC-MC sample, confirming the absence of bias in the BF measurement.

\section{Systematic uncertainties}
\label{sys}
In the following, the main contributions to the systematic uncertainties, as summarized in Table~\ref{tab:sys}, are discussed. In the ratio $\mathcal{ R}$ of BFs, defined in Eq.~\ref{eq:R}, the systematic uncertainties for the tracking efficiency of the charged tracks and for $N_{\psi (2S) }$ cancel.

The relative uncertainty due to the finite size (750000 events) of the exclusive MC sample used for the calculation of efficiency is calculated by binomial error calculation, resulting in 0.2\%.
\begin{table}[htbp] 
  \caption{ Relative systematic uncertainties in the BF measurement.  }
  \begin{center}
\begin{tabular}{|l|l|}\hline\hline
Source & Systematic uncertainty (\%)\\
\hline
Event selection   &1.5\\
MC statistics & 0.2\\
Background  &0.2\\
$\mathcal{B}_{\mu^+\mu^-}$  &0.5\\
\hline
Total& 1.6\\
\hline
\hline
\end{tabular}
\end{center}
\label{tab:sys}
\end{table}
For all the sources related to the event selection, the systematic uncertainty is calculated from the maximum difference between the $\mathcal{B}_{K^+K^-}$ values obtained by changing the selection and the nominal one, with both the efficiency and the signal yield being affected by the event selections.
This selection is used to tag the $J/\psi$ and reject the background, mostly the non-$\pi^+\pi^- J/\psi$ background.
In detail the \rm{RM} window is modified from about 5$\sigma$ to 4$\sigma$ and we calculate the difference in the $\mathcal{B}_{K^+K^-}$.
The relative uncertainty is 0.30\%. The limits for $\cos \theta$ is tightened to be $\pm 0.7$. The corresponding uncertainty is negligible, being less than  $2\times 10^{-5}$.
We change the selections on $\cos\theta_{\pi\pi}$ and $\cos \theta_{KK}$ to estimate the related systematic uncertainty. 
In the first case we change the upper limit on $\cos\theta_{\pi\pi}$  from 0.5 to 0.6, and in the second case we change the upper limit on $\cos \theta_{K^+K^-}$ from $-0.95$ to $-0.9$. In either case, the systematic uncertainty is negligible, being less than $5\times10^{-5}$.
We modify the selection lowering the kaon momenta cut to 1.0 GeV/c. The uncertainty is negligible, being around  $2\times 10^{-5}$.
The normalized net momentum selection is changed by $\pm 0.05$, keeping the largest difference in $\mathcal{B}$ as systematic uncertainty. It is found to be 1.00\%.
The $E_{dep}/p$ upper limit is changed to 0.7 and 0.9, taking the larger difference as the systematic uncertainty with a value of 0.7\% .
The $E_{EMC}$ window is modified to [0.35, 2.45] and [0.25, 2.55]~GeV. The systematic uncertainty is found to be 0.5\%.
The $X_{\rm{vis}}$ range, which defines the signal range after fitting, is changed from the chosen asymmetric window, fixed to [0.98, 1.01], due to the dimuon peak in the higher $X_{\rm{vis}}$ values range, to [-4$\sigma$, +2$\sigma$], where the $\sigma$ is obtained from the full width at half maximum of the signal distribution. The systematics is 0.6\% 
The total relative systematic uncertainty on the event selection is 1.50\%.
The uncertainty due to the peaking background fraction is obtained as the maximum variation in the $\mathcal{B}_{K^+K^-}$ changing the background fraction by $\pm 1 \sigma$. The uncertainty is evaluated to be 0.2\%.
The systematic uncertainty for $\mathcal{B}_{\mu^+\mu^-}$ is quoted from the PDG as 0.5\%~\cite{pdg}.
The uncertainty due to the fit procedure is estimated by adding an additional flat background function and results in being negligible.
Furthermore a study of pseudoexperiments is performed to evaluate additional systematics in the fitting procedure, and no fit bias is found.
The sum in quadrature of all the contributions gives a relative total systematic uncertainty of 1.6\%, corresponding to $ 0.050\times 10^{-4}$.

\section{Summary}
\label{summary}
We have performed a precision measurement of the BF of the decay of $J/\psi\to K^+K^-$ to be $\mathcal{B}_{K^+K^-}=(3.072\pm 0.023({\rm stat.})\pm 0.050({\rm syst.}))\times 10^{-4}$, still dominated by the systematic uncertainty.
This is an improvement by more than a factor of three over the previous measurement of $(2.86\pm 0.21)\times 10^{-4}$\cite{cleoc}, obtained using the CLEO-c data, that is up to now the PDG value, with central value in agreement with that BF result.
The most precise measurement of the $\mathcal{B}_{K^+K^-}$ in direct reaction, which takes into account the interference effect, is the BaBar one~\cite{babar}, in which they gave an indication of the positive sign of the relative phase.
Due to their large uncertainties, our result is in agreement with both $\mathcal{B}_{K^+K^-}^{\rm{BaBar+}}=(3.22\pm 0.20\pm 0.12)\times 10^{-4}$ with positive phase within less than 1$\sigma$ and $\mathcal{B}_{K^+K^-}^{\rm{BaBar-}}=(3.50\pm 0.20\pm 0.12)\times 10^{-4}$ with negative phase within about 2$\sigma$.
Our result represents the first step in the path to determine the sign of the relative phase between strong and electromagnetic amplitudes in this channel. A measurement in the $e^+e^- \rightarrow K^+K^-$ process with improved precision will complete the picture.

\section{Acknowledgement}

The BESIII Collaboration thanks the staff of BEPCII and the IHEP computing center for their strong support. This work is supported in part by National Key R\&D Program of China under Contracts Nos. 2020YFA0406300, 2020YFA0406400; National Natural Science Foundation of China (NSFC) under Contracts Nos. 11635010, 11735014, 11835012, 11935015, 11935016, 11935018, 11961141012, 12025502, 12035009, 12035013, 12061131003, 12192260, 12192261, 12192262, 12192263, 12192264, 12192265, 12221005, 12225509, 12235017; the Chinese Academy of Sciences (CAS) Large-Scale Scientific Facility Program; the CAS Center for Excellence in Particle Physics (CCEPP); Joint Large-Scale Scientific Facility Funds of the NSFC and CAS under Contract No. U1832207; CAS Key Research Program of Frontier Sciences under Contracts Nos. QYZDJ-SSW-SLH003, QYZDJ-SSW-SLH040; 100 Talents Program of CAS; The Institute of Nuclear and Particle Physics (INPAC) and Shanghai Key Laboratory for Particle Physics and Cosmology; European Union's Horizon 2020 research and innovation programme under Marie Sklodowska-Curie grant agreement under Contract No. 894790; German Research Foundation DFG under Contracts Nos. 455635585, Collaborative Research Center CRC 1044, FOR5327, GRK 2149; Istituto Nazionale di Fisica Nucleare, Italy; Ministry of Development of Turkey under Contract No. DPT2006K-120470; National Research Foundation of Korea under Contract No. NRF-2022R1A2C1092335; National Science and Technology fund of Mongolia; National Science Research and Innovation Fund (NSRF) via the Program Management Unit for Human Resources \& Institutional Development, Research and Innovation of Thailand under Contract No. B16F640076; Polish National Science Centre under Contract No. 2019/35/O/ST2/02907; The Swedish Research Council; U. S. Department of Energy under Contract No. DE-FG02-05ER41374


\bibliographystyle{apsrev4-2} 
\bibliography{draft_bb.bib}

\end{document}